\documentclass[10pt,twocolumn]{article}
\usepackage[utf8]{inputenc}
\usepackage[T1]{fontenc}
\usepackage{lmodern}
\usepackage{amsmath,amssymb}
\usepackage{graphicx}
\usepackage{booktabs}
\usepackage{hyperref}
\usepackage{enumitem}
\usepackage{multicol}
\usepackage[margin=0.75in,columnsep=0.25in]{geometry}
\usepackage{pdflscape}
\usepackage{rotating}

\title{VEHRON: A Configuration-Driven BEV Simulation Framework for Subsystem-Level Studies}
\author{
Subramanyam Natarajan\\
\texttt{sn@chmod.in}
}
\date{26 April 2026}

\begin{document}
\twocolumn[
\maketitle
\begin{@twocolumnfalse}
\begin{abstract}
In practical early-stage battery-electric vehicle studies, analysis workflows
may become fragmented across spreadsheets, notebooks, and project-specific
scripts, making reuse, audit, and extension harder. VEHRON is an open-source
Python framework for a deterministic, traceable workflow built around
prescribed-speed longitudinal simulation of battery-electric vehicles using validated YAML configuration,
packaged drive-cycle resources, interchangeable subsystem models, and
auditable case outputs. VEHRON currently runs as a command-line workflow in
which a vehicle definition and a testcase definition are combined to execute a
simulation, emit a flat time series, and write a case package containing
copied inputs, resolved configuration, summary metadata, and standard plots.
Architecturally, VEHRON is organized around a small simulation engine, a
shared state bus, a registry of model selections, schema-based configuration
loading, and extension points for custom battery and HVAC models loaded from
external Python files. VEHRON currently focuses on battery-electric
longitudinal simulation with low-order battery, thermal, auxiliary-load, and
HVAC models. This paper explains how VEHRON is structured, how it is used,
which models it implements, and where its present limits lie. Source code is available at
\url{https://github.com/vehron-dev/vehron}~\cite{vehronrepo}, with archived
release metadata recorded under DOI \url{https://doi.org/10.5281/zenodo.19820111}.
\end{abstract}
\end{@twocolumnfalse}
]


\section{Introduction}
BEV energy studies in engineering and research settings are frequently done
with a mix of scripts, notebooks, and one-off tools assembled for a single
project. The problem is not the analysis — it is that assumptions live in
code comments, outputs go wherever is convenient, and the next person who
needs to repeat or extend the study starts from scratch. That is a workflow
problem, and it compounds over time.

VEHRON (VEHicle Research and Optimisation Network)~\cite{vehronrepo} is built
to fix that workflow for one specific slice of the problem: battery-electric
vehicle longitudinal simulation. The design is deliberately narrow. You define
a vehicle and a testcase in YAML. You run from the CLI. You get a structured
output package with copied inputs, timeseries, plots, and a summary JSON. The
same vehicle file runs against any testcase without modification.

This paper describes the software as it exists in the repository: package
structure, public interfaces, configuration model, implemented models,
verification practices, and current scope.

VEHRON's specific contribution is the combination of three design decisions
working together:
(i)~YAML-driven configuration, where every vehicle parameter, model
selection, and mission condition is explicitly stated and version-controllable
— study assumptions are auditable without reading source code;
(ii)~subsystem slot separation, where battery and HVAC models are swappable
by pointing a YAML field at a local Python file, so a proprietary or
higher-fidelity model can run inside the same vehicle-level context without
touching the engine; and
(iii)~deterministic case packaging as a first-class output, where each run
writes a self-contained artifact — copied inputs, resolved config, timeseries,
summary metadata, plots — making reruns and comparisons straightforward.

\subsection{Relationship to Existing Tools}
Several established tools address vehicle energy simulation.
ADVISOR~\cite{advisor2002} and Autonomie~\cite{autonomie2011} are established
MATLAB/Simulink-oriented vehicle-system simulation environments with broad
powertrain coverage. FASTSim~\cite{fastsim2018} from NREL is Python-based and
well suited to fleet-level consumption analysis. OpenModelica~\cite{openmodelica2020}
and other Modelica environments support high-fidelity multi-physics modelling.
Commercial tools like GT-SUITE~\cite{gtsuite_web} and IPG CarMaker~\cite{ipg_carmaker_web}
offer broad vehicle-simulation capability as closed commercial products. VEHRON
differs in emphasizing a lightweight CLI workflow, explicit subsystem slots,
and deterministic case-package outputs.

VEHRON occupies a narrower space: a researcher or engineer who needs to put a
custom battery model, HVAC strategy, or control logic into a vehicle-level
simulation context — without building the surrounding infrastructure from
scratch, and without a commercial licence or a MATLAB dependency. VEHRON's
contribution is its workflow: a clean way to run subsystem studies inside a
vehicle-level simulation.

\section{Motivation and Problem Context}
VEHRON addresses everyday BEV study tasks that are easy to state and tedious
to structure well. A user wants to compare duty cycles
for a candidate BEV. Or inspect SoC trajectories under different ambient
conditions. Or see what happens to range when auxiliary loads increase. Or hand
a validated battery model to a simulation without rebuilding everything around
it. The difficulty is not the equations; it is keeping the workflow consistent
across runs.

Three design goals follow directly from this. Vehicle hardware and mission
conditions live in separate YAML files, so the same vehicle runs across any
number of testcases without duplication. Every run writes a deterministic case
package, so you can rerun or compare months later without reconstructing what
you did. Subsystem models sit behind explicit slot interfaces, so replacing or
upgrading a model does not touch the engine.

VEHRON in its current form is a forward-time, CLI-driven, YAML-configured BEV
workflow with low-order energy and thermal submodels. HEV, PHEV, and FCEV
paths are on the roadmap, but this paper focuses on the BEV path.

\section{Intended Use Cases}
The core idea is that you should only need to build the subsystem you are
actually studying. VEHRON supplies the rest: longitudinal dynamics, energy
flow, drive-cycle execution, thermal coupling, case packaging.

An HVAC team building a new cabin-conditioning strategy implements one Python
class from \texttt{HvacModelBase}, points the YAML at it, and runs the same
model across multiple cycles and ambient conditions. The battery, motor,
route, and output packaging are already there.

A battery team with a validated pack model drops it into the battery slot via
\texttt{BatteryModelBase}, points the YAML at their local Python file, and
runs WLTP or custom cycles before physical testing. The proprietary code stays
outside the repository.

A researcher doing a parametric sweep stays entirely at the YAML and CLI level
— vary mass, auxiliary loads, HVAC COP, run, compare. No engine modifications
needed. Each run produces a self-contained artifact: timeseries, summary,
plots, and copied inputs. That makes archiving and comparison straightforward
in a way that notebook-based workflows typically are not.

\subsection{External Model Slots}
Battery and HVAC models can be loaded from local Python files at runtime.
This is the mechanism that lets a proprietary subsystem model run inside
VEHRON without being committed to a public repository. The simulation
infrastructure stays open; the controlled implementation stays private.

\section{Software Design and Architecture}
VEHRON is implemented as a Python package under \texttt{src/vehron}. The core
architectural components visible in the repository are:
\begin{itemize}[leftmargin=1.5em]
\item \texttt{engine.py}, which instantiates modules and runs the simulation
loop;
\item \texttt{state.py}, which defines the shared simulation state passed
between modules;
\item \texttt{loader.py}, which loads vehicle and testcase YAML files and
applies schema-based validation and boundary conversions;
\item \texttt{registry.py}, which maps configuration model selections to
runtime classes;
\item \texttt{routes.py}, which handles route and drive-cycle related input;
\item \texttt{runner.py}, which provides the command-line entry points;
\item \texttt{post/}, which generates reports, time-series exports, and plots;
\item \texttt{schemas/}, which contains configuration schemas for vehicles,
testcases, and module parameters;
\item \texttt{modules/}, which groups subsystem models by physical domain.
\end{itemize}

The core principle is separation of concerns. \texttt{SimEngine} owns
orchestration and time stepping — it contains no physics. Subsystem physics
live in dedicated modules. Runtime state is held in a shared \texttt{SimState}
object that acts as the bus between modules: modules read from it, compute
outputs, and write back through the engine's update path. Configuration comes
in through validated YAML, so experiment setup stays visible and editable
outside the loop.

The BEV runtime flow is: load and validate two YAML files, apply boundary-unit
conversions at the loader edge, instantiate subsystem modules, advance the
master clock, collect outputs into a flat timeseries, write the case package.

The runtime uses fixed-step multi-rate execution. Master timestep is $\Delta t$
(0.1\,s by default). If a module declares a rate divisor $d$, its effective
update interval is
\begin{equation}
  \Delta t_{\mathrm{eff}} = d\,\Delta t.
  \label{eq:multirate}
\end{equation}
Fast domains — longitudinal dynamics, motor, inverter, regen — run every
master step. Slower domains — battery electrical, HVAC, thermal trends — run
less often. To avoid distortion from sampling a slow module on a single
instantaneous input, VEHRON accumulates intermediate signals across master
steps and flushes the accumulator just before the slow module steps.

VEHRON v1 is a BEV-focused package. This paper stays within that boundary.
Hybrid-electric, fuel-cell, and richer route inputs (elevation, GPS replay)
are planned but not yet part of the supported surface.

\section{Implemented Models}
This section describes the reference model stack used by VEHRON's BEV path.
Together, these models provide the vehicle context in which battery, thermal,
and auxiliary-load behaviour is studied. Anything here can be replaced via the
extension interfaces without touching the engine.

The baseline formulations follow standard vehicle-dynamics and
electrified-vehicle texts~\cite{gillespie1992fvd,rajamani2012vdc,liu2013hev}.

\subsection{Longitudinal Dynamics}
The longitudinal vehicle model uses prescribed-speed motion. The route or
drive-cycle input provides the speed history \(v(t)\), and VEHRON derives the
corresponding acceleration, wheel force, throttle/brake demand, torque, and
power from that imposed motion. The required wheel force is resolved from the
standard road-load decomposition used in vehicle-dynamics and
electrified-vehicle texts~\cite{gillespie1992fvd,rajamani2012vdc,liu2013hev}:
\begin{equation}
  F_{\mathrm{wheel,req}} = ma + F_{\mathrm{aero}} + F_{\mathrm{roll}} + F_{\mathrm{grade}},
\end{equation}
\begin{equation}
  a_k = \frac{v_{k+1} - v_k}{\Delta t}.
\end{equation}
Distance is then advanced kinematically as
\begin{align}
  x_{k+1} &= x_k + \frac{v_k + v_{k+1}}{2}\,\Delta t.
\end{align}
The implemented road-load terms are
\begin{align}
  F_{\mathrm{aero}} &= \tfrac{1}{2}\rho C_d A\,v_{\mathrm{rel}}^2, \\
  F_{\mathrm{roll}} &= mg\,C_{rr}\cos(\theta)\,(1 + 0.01\,v), \\
  F_{\mathrm{grade}} &= mg\sin(\theta),
\end{align}
with derived traction and braking demands obtained by comparing the required
wheel force against the configured drive and brake force limits. Here $\rho$
is air density, $C_d$ is drag coefficient, $A$ is frontal area,
$v_{\mathrm{rel}}$ is relative air speed, $C_{rr}$ is rolling-resistance
coefficient, and $\theta$ is road grade angle.
The multiplicative $(1 + 0.01\,v)$ factor is a VEHRON low-order
rolling-resistance correction used to modestly increase tyre drag with speed in
the prescribed-speed runtime; it should be read as an implementation
assumption rather than a separately identified tyre law.

\subsection{Reducer, Motor, and Inverter}
The reducer maps wheel-side quantities to the motor shaft through a fixed total
ratio $r_{\mathrm{tot}} = r_{\mathrm{primary}}\,r_{\mathrm{secondary}}$.
Wheel speed and motor speed are related by
\begin{align}
  \omega_{\mathrm{wheel}} &= \frac{v}{r_{\mathrm{wheel}}}, &
  \omega_{\mathrm{motor}} &= r_{\mathrm{tot}}\,\omega_{\mathrm{wheel}}.
\end{align}
VEHRON's default \texttt{analytical} motor model is an
envelope-based EV traction motor model with separate motoring and regenerative
limits. For positive torque request, the admissible torque is
\begin{equation}
  T_{\mathrm{lim,mot}}(\omega) =
  \begin{cases}
    T_{\mathrm{pk}}, & \omega \le \omega_{\mathrm{base}},\\[4pt]
    \min\!\left(T_{\mathrm{pk}}, \dfrac{P_{\mathrm{pk}}}{\omega}\right),
      & \omega > \omega_{\mathrm{base}},
  \end{cases}
\end{equation}
where $T_{\mathrm{pk}}$ is the configured peak torque,
$P_{\mathrm{pk}}$ is the configured peak power, and
$\omega_{\mathrm{base}}$ is either provided directly as
\texttt{motor.base\_speed\_rpm} or derived from
$\omega_{\mathrm{base}} = P_{\mathrm{pk}}/T_{\mathrm{pk}}$ and then clipped by
the configured maximum motor speed. The commanded positive torque is then
clamped to the interval $[0,\,T_{\mathrm{lim,mot}}]$.

For regenerative operation, the model uses analogous limits with optional
separate ceiling parameters:
\begin{equation}
  T_{\mathrm{lim,regen}}(\omega) =
  \begin{cases}
    T_{\mathrm{regen,max}}, & \omega \le \omega_{\mathrm{base}},\\[4pt]
    \min\!\left(T_{\mathrm{regen,max}},
    \dfrac{P_{\mathrm{regen,max}}}{\omega}\right),
      & \omega > \omega_{\mathrm{base}},
  \end{cases}
\end{equation}
and the requested negative torque is clamped to
$[-T_{\mathrm{lim,regen}},\,0]$. If the user does not provide
\texttt{motor.max\_regen\_torque\_nm} or
\texttt{motor.max\_regen\_power\_kw}, the model falls back to the motoring
limits. This keeps the model usable when separate regen ceiling data are unavailable.

Mechanical power is computed in the usual way:
\begin{equation}
  P_{\mathrm{mech}} = T_{\mathrm{motor}}\,\omega_{\mathrm{motor}}.
\end{equation}
The scalar analytical efficiency model is then
\begin{equation}
  \eta_{\mathrm{motor}} = \mathrm{clip}\!\left(\eta_{\mathrm{base}}
    - 0.06(1{-}\lambda_T) - 0.03\lambda_\omega,\;
    \eta_{\min},\; \eta_{\max}\right),
\end{equation}
where $\lambda_T = |T_{\mathrm{motor}}|/T_{\mathrm{pk}}$ and
$\lambda_\omega = \omega_{\mathrm{motor}}/\omega_{\max}$ are capped at unity.
The parameters $\eta_{\mathrm{base}}$, $\eta_{\min}$, and $\eta_{\max}$ are
read from YAML as \texttt{base\_efficiency}, \texttt{min\_efficiency}, and
\texttt{max\_efficiency}. An optional \texttt{regen\_efficiency} can override
this expression on the regenerative branch. These efficiency terms are
calibration parameters for low-order studies, not identification of a
machine-loss model. In practical use, the torque, power, and speed ceilings
should come from motor or drivetrain specification data, while
\texttt{base\_speed\_rpm}, \texttt{base\_efficiency}, and the efficiency bounds
should be chosen to match the intended torque envelope and overall energy use for the study.
The functional form is VEHRON's own compact scalar approximation for an EV
traction-motor envelope, informed by the general efficiency-shaping discussion
in the traction-motor literature rather than copied from a single published
loss model~\cite{agrawal2024traction}.
At zero torque, this scalar expression remains bounded by the same clipped
speed-dependent form and therefore acts as a lightweight coasting-side
efficiency approximation rather than a separate idle-loss model.
As a VEHRON benchmark sensitivity, reducing \texttt{base\_efficiency} by two
percentage points in the flat-highway sedan benchmark increases net energy
consumption by approximately \(120.4\) Wh (\(2.24\%\)). This indicates that
users should treat this parameter as a calibration input requiring pack-
specific or map-based data where energy accuracy is important.

For positive motoring power, the electrical demand is
$P_{\mathrm{drive}} = P_{\mathrm{mech}}/\eta_{\mathrm{motor}}$. Negative motor
mechanical power is not itself written back as battery charging power by this
module; in VEHRON, regenerative energy recovery is handled by the
blended-braking module described below. The inverter
then applies a fixed efficiency:
\begin{equation}
  P_{\mathrm{drive,dc}} = \frac{P_{\mathrm{drive}}}{\eta_{\mathrm{inv}}}.
\end{equation}

The second in-repo motor option, \texttt{efficiency\_map}, reuses the same
torque-speed envelope and then replaces the scalar efficiency estimate with
CSV-backed lookup when a map file is present. The implementation uses
nearest-neighbour search in the speed-torque plane. Map smoothness therefore
depends on the supplied grid density.

\subsection{Regenerative Braking}
\label{sec:regen}
The blended braking model routes a configurable fraction of the available
kinetic braking opportunity to the regenerative path. In prescribed-speed
simulation the instantaneous wheel power is derived from imposed motion, so the
kinetic opportunity is taken directly from the wheel-side power signal:
\begin{equation}
  P_{\mathrm{kinetic}} = \max\!\left(-P_{\mathrm{wheel}},\;0\right).
\end{equation}
The recoverable power, before hardware ceiling, is scaled by a blend factor
$\beta \in [0,1]$ that represents the fraction of braking events actually
routed to the motor rather than friction:
\begin{equation}
  P_{\mathrm{hw}} = \min\!\left(\beta\,P_{\mathrm{kinetic}},\;
                               P_{\mathrm{regen,max}}\right),
\end{equation}
and the electrical power returned to the battery is
\begin{equation}
  P_{\mathrm{regen}} = \eta_{\mathrm{regen}}\,P_{\mathrm{hw}}.
\end{equation}
Here $P_{\mathrm{regen,max}}$ is the configured hardware regen ceiling,
$\eta_{\mathrm{regen}}$ is the motor-path regen efficiency read from YAML, and
$\beta$ is the \texttt{regen\_blend\_factor} parameter. Regen is suppressed at
speeds below 0.5~m/s, when no brake demand is present, or when SoC is at or
above the runtime hard upper limit. The 0.5~m/s cutoff is a VEHRON numerical
guard chosen to avoid spurious low-speed regen chatter in prescribed-speed
runs; it is not presented as a regulatory or OEM threshold.

The parameter $\beta$ is a calibration scalar, not a hardware specification.
It encodes the aggregate fraction of kinetic braking opportunity that a given
vehicle-strategy combination routes to the motor during a drive cycle. In an
ideal one-pedal vehicle with no friction-only stops, $\beta = 1$; in practice
it is below unity because friction blending occurs at low speed, during
emergency stops, and wherever the brake blend strategy limits regen contribution~\cite{szumska2025regen}.
The correct value for a given vehicle on a given cycle is
determined by energy-balance calibration against a published consumption figure,
as described in Section~\ref{sec:refvehicle}.

\subsection{Battery Models}
The baseline \texttt{rint} battery model resolves the net battery-side power
demand as traction plus HVAC plus auxiliaries minus regenerative recovery and
any configured external charging contribution. Using nominal voltage
$V_{\mathrm{nom}}$, the ideal current is
\begin{equation}
  I_{\mathrm{ideal}} = \frac{P_{\mathrm{net}}}{V_{\mathrm{nom}}},
\end{equation}
which is then clamped by charge and discharge C-rate limits. Terminal voltage
is
\begin{equation}
  V_{\mathrm{batt}} = V_{\mathrm{nom}} - I_{\mathrm{batt}}\,R_{\mathrm{int}},
\end{equation}
and SoC integration is
\begin{equation}
  \mathrm{SoC}_{k+1} = \mathrm{clip}\!\left(
    \mathrm{SoC}_k - \frac{I_{\mathrm{batt}}\,\Delta t}{Q_{\mathrm{Ah}}\cdot 3600},\;
    \mathrm{SoC}_{\min},\;
    \mathrm{SoC}_{\max}\right).
\end{equation}

The higher-fidelity equivalent-circuit battery implementation includes a
two-RC ECM family, as described in representative battery-modeling literature
such as Hu et al.~\cite{hu2012ecm} and Su et al.~\cite{su2019ecm}, with
terminal voltage represented as
\begin{equation}
  V_{\mathrm{term}} = V_{\mathrm{ocv}} - IR_0 - V_{\mathrm{rc1}} - V_{\mathrm{rc2}},
\end{equation}
where $V_{\mathrm{ocv}}$ is a shaped open-circuit-voltage estimate from SoC.
Each RC branch evolves using the exponential discrete-time update:
\begin{align}
  V_{\mathrm{rc},i}^{k+1} &= \alpha_i V_{\mathrm{rc},i}^{k}
    + (1-\alpha_i)\,I_k\,R_i, \\
  \alpha_i &= \exp\!\left(-\frac{\Delta t}{R_i C_i}\right).
\end{align}
The ECM therefore adds transient sag and recovery behavior while remaining
fast enough for vehicle-level simulation.

\subsection{External Charging Model}
VEHRON includes an in-repository external charging controller for AC
charging. Charging availability and limits are split across configuration in a
way that mirrors the distinction between capability and use-case selection.
Vehicle YAML can declare charger-side capability fields such as
\texttt{ac\_power\_limit\_kw}, charge efficiency, charge-current limits, target
voltage, and temperature guard bands, while testcase YAML selects whether
charging is enabled for a run and defines the time window during which the
vehicle is considered plugged in.

The \texttt{ac\_basic} implementation is a battery-side request model
with a simple constant-power / constant-voltage style transition. While the
plug-in window is active, the model first requests an approximately constant
battery-side charging power,
\begin{equation}
  P_{\mathrm{chg,req}} = -\eta_{\mathrm{ac}} P_{\mathrm{ac,max}},
\end{equation}
subject to an optional maximum battery current. If battery terminal voltage
reaches the configured target voltage, the controller switches to a crude
constant-voltage taper based on a resistance-derived open-circuit estimate:
\begin{align}
  V_{\mathrm{ocv,est}} &= V_{\mathrm{batt}} + I_{\mathrm{batt}}R_{\mathrm{chg}}, \\
  I_{\mathrm{cv}} &= \max\!\left(
    \frac{V_{\mathrm{target}} - V_{\mathrm{ocv,est}}}{R_{\mathrm{chg}}},\;0\right).
\end{align}
Here $R_{\mathrm{chg}}$ is a user-supplied charging-path resistance parameter.
If the tapered current falls below the configured termination threshold, the
charger enters a done state and requests zero power. The model also blocks
charging if the battery temperature is outside optional minimum or maximum
charging bounds. This charging path is a runtime request model for cycle-level charging studies with user-supplied charger parameters. Correspondingly, parameters such as charge efficiency,
charge resistance, termination current, and target voltage are user-supplied
control and calibration parameters, not quantities estimated internally by the
software.

\subsection{Auxiliary Electrical Loads}
The present auxiliary-load model is intentionally simple. It sums configured
electrical parasitics:
\begin{equation}
  P_{\mathrm{aux}} = P_{\mathrm{headlights}} + P_{\mathrm{adas}}
                   + P_{\mathrm{infotainment}} + P_{\mathrm{steering}}.
\end{equation}
This gives users a documented and auditable way to vary auxiliary demand
through YAML configuration.

\subsection{Cabin HVAC and Cabin Thermal Model}
The cabin HVAC model follows the low-order vehicle-cabin thermal-model family
used in comparative energy studies, as discussed for example by Marcos et
al.~\cite{marcos2014cabin}, Torregrosa-Jaime et al.~\cite{torregrosa2015cabin},
Marshall et al.~\cite{noreen2019thermal}, and reviewed in~\cite{cabreview2024}.
The cabin is treated as a single lumped thermal mass, with passive and active
heat terms combined as
\begin{equation}
  Q_{\mathrm{passive}} = Q_{\mathrm{envelope}} + Q_{\mathrm{solar}}
    + Q_{\mathrm{ventilation}} + Q_{\mathrm{occupants}},
\end{equation}
and net cabin heat
\begin{equation}
  Q_{\mathrm{net}} = Q_{\mathrm{passive}} + Q_{\mathrm{hvac}}.
\end{equation}
The implemented terms are
\begin{align}
  Q_{\mathrm{envelope}} &= UA_{\mathrm{tot}}(T_{\mathrm{amb}} - T_{\mathrm{cabin}}), \\
  UA_{\mathrm{tot}} &= UA_{\mathrm{body}} + k_v v, \\
  Q_{\mathrm{solar}} &= G_{\mathrm{solar}}\,A_{\mathrm{glass}}\,\tau_{\mathrm{solar}}, \\
  Q_{\mathrm{vent}} &= \dot{m}_{\mathrm{air}}\,c_p(T_{\mathrm{amb}} - T_{\mathrm{cabin}}), \\
  Q_{\mathrm{occ}} &= N_{\mathrm{occ}}\,q_{\mathrm{occ}},
\end{align}
and cabin temperature evolves as
\begin{equation}
  T_{\mathrm{cabin},k+1} = T_{\mathrm{cabin},k}
    + \frac{Q_{\mathrm{net}}\,\Delta t}{C_{\mathrm{tot}}}.
\end{equation}
Here $UA_{\mathrm{tot}}$ denotes the overall cabin heat-transfer conductance,
$A_{\mathrm{glass}}$ the glazed area, $\tau_{\mathrm{solar}}$ an effective
solar transmittance, $\dot{m}_{\mathrm{air}}$ the ventilation air mass flow,
$c_p$ the air specific heat, and $C_{\mathrm{tot}}$ the effective cabin
thermal capacitance.
Cooling and heating thermal requests are clipped by the rated HVAC thermal
power. Electrical HVAC demand is then computed from the relevant coefficient
of performance:
\begin{equation}
  P_{\mathrm{hvac}} = \begin{cases}
    \dfrac{|Q_{\mathrm{hvac}}|}{\mathrm{COP}_{\mathrm{cooling}}}, & Q_{\mathrm{hvac}} < 0, \\[6pt]
    \dfrac{Q_{\mathrm{hvac}}}{\mathrm{COP}_{\mathrm{heating}}},   & Q_{\mathrm{hvac}} > 0, \\[6pt]
    0, & Q_{\mathrm{hvac}} = 0.
  \end{cases}
\end{equation}

\subsection{Thermal Trend Models}
VEHRON currently uses three low-order thermal trend models in the active BEV
path. The battery thermal model applies first-order ambient relaxation plus a
current-dependent loss term:
\begin{equation}
  T_{\mathrm{batt},k+1} = T_{\mathrm{batt},k}
    + \frac{T_{\mathrm{amb}} - T_{\mathrm{batt}}}{\tau_{\mathrm{batt}}}\,\Delta t
    + \beta_{\mathrm{batt}}\,Q_{\mathrm{loss}}\,\Delta t,
\end{equation}
where $Q_{\mathrm{loss}} = 3|I_{\mathrm{batt}}|$. The coefficient 3 is a VEHRON
scaling constant used to couple battery current magnitude to a low-order
thermal trend; it is an implementation tuning choice rather than a directly
measured heat-generation law.
The motor thermal model uses the same first-order structure with loss power
derived from shaft power and motor inefficiency:
\begin{align}
  Q_{\mathrm{loss,m}} &= |T_{\mathrm{m}}\,\omega_{\mathrm{m}}|\,(1-\eta_{\mathrm{m}}), \\
  T_{\mathrm{m},k+1} &= T_{\mathrm{m},k}
    + \frac{T_{\mathrm{amb}} - T_{\mathrm{m}}}{\tau_{\mathrm{m}}}\,\Delta t
    + \beta_{\mathrm{m}}\,Q_{\mathrm{loss,m}}\,\Delta t,
\end{align}
where the subscript $\mathrm{m}$ denotes motor quantities.
The coolant loop then uses a simple first-order relaxation toward the mean of
the battery and motor temperatures:
\begin{align}
  T_{\mathrm{target}} &= \frac{T_{\mathrm{batt}} + T_{\mathrm{motor}}}{2}, \\
  T_{\mathrm{coolant},k+1} &= T_{\mathrm{coolant},k}
    + \frac{T_{\mathrm{target}} - T_{\mathrm{coolant}}}{\tau_{\mathrm{coolant}}}\,\Delta t.
\end{align}

\section{Implementation Details}
The package metadata, CLI entry points, schemas, and bundled resources
described in this section correspond to the VEHRON v0.2.2 release~\cite{vehronrepo}.
VEHRON is distributed as a Python package with \texttt{setuptools}-based
build metadata in \texttt{pyproject.toml}. The package declares Python
\texttt{>=3.10} and depends on \texttt{numpy}, \texttt{pydantic},
\texttt{pyyaml}, \texttt{click}, and \texttt{matplotlib}. The command-line
entry point is installed as \texttt{vehron} via
\texttt{vehron.runner:cli}.

The configuration model is central to the implementation. Vehicle YAML files
describe hardware, subsystem availability, and model selection, while testcase
YAML files describe route mode, environmental conditions, payload, run-time
charger selection, and simulation settings. The schema layer includes explicit
fields for payload assumptions such as passenger count, passenger mass, and
cargo mass. It distinguishes charger capability from charger use:
vehicle-side configuration can define charging limits and control parameters,
while testcase-side configuration can enable or disable charging for a run and
define the active plug-in window.

Subsystem implementations are grouped by domain. The repository includes
longitudinal dynamics, battery models, HVAC models, thermal trend
models, auxiliary loads, charging models, and post-processing tools. The
package also contains resource helpers that expose packaged archetypes,
packaged testcases, and bundled drive-cycle CSV data, allowing examples to run
with packaged resources available from the start.

The command-line interface exposes at least four user-facing commands:
\texttt{vehron init}, \texttt{vehron run}, \texttt{vehron list-examples},
and \texttt{vehron run-example}. The case-directory workflow begins with
\texttt{vehron init}, which writes a \texttt{.vehron-case} marker,
\texttt{README.md}, \texttt{vehicle.yaml}, \texttt{testcase.yaml}, and an
empty \texttt{output/} directory. The main run path then prints a short
specification sheet, executes the simulation, reports selected summary values
such as steps, distance, final SoC, and total energy, and writes each run
under the case directory's \texttt{output/} tree.

\section{Core Capabilities}
The capabilities listed in this section are part of the VEHRON v0.2.2
release~\cite{vehronrepo}.
\subsection{Configuration-Driven BEV Simulation}
The primary workflow is: provide a vehicle YAML and a testcase YAML, run
\texttt{vehron run}, get a case package. That is the whole public interface.
No code changes required between studies.

\subsection{Drive-Cycle and Parametric Route Inputs}
Route input is either a parametric definition or a speed-trace CSV with columns
\texttt{time\_s,speed\_kmh}. Several drive cycles are packaged with the
repository under \texttt{src/vehron/data/drive\_cycles/}, including the full
1800\,s WLTP Class~3b trace used in Section~\ref{sec:refvehicle}.

\subsection{Modular Subsystem Models}
The active BEV chain covers prescribed-speed longitudinal dynamics, fixed-ratio
reduction, analytical and map-based motor models, inverter, blended regen,
battery (Rint and 2RC ECM), auxiliary loads, AC charging, HVAC, and low-order
thermal trends. Model selection — motor type, charging on/off, battery model —
is done in YAML, not in code.

\subsection{External Battery and HVAC Slots}
Battery and HVAC models load from a local Python file at runtime via
\texttt{battery.external\_module\_path} and \texttt{hvac.external\_module\_path}.
The repository includes stub examples. The outer simulation infrastructure
stays open; the implementation stays private.

\subsection{Case Packaging}
Each run writes under \texttt{<case\_dir>/output/<case\_name>/}:
\texttt{summary.json} (including the full road-load energy breakdown),
\texttt{timeseries.csv}, copied and resolved input YAMLs, a \texttt{README.md},
and plots. The summary JSON is machine-readable, so runs can be compared
directly without manual cleanup.

\subsection{Packaged Examples}
Archetypes, testcases, and drive-cycle CSVs ship with the package. The six
reference vehicle cases from Section~\ref{sec:refvehicle} are included and
runnable with a single \texttt{vehron run} command.

\section{Example Workflow and Case Artifact}
This workflow description refers to the VEHRON v0.2.2 release~\cite{vehronrepo}.
Install from a checkout and run the baseline sedan:
\begin{small}
\begin{verbatim}
python3 -m venv .venv && source .venv/bin/activate
pip install .
vehron init veh-case-1
vehron run {-}{-}case veh-case-1
\end{verbatim}
\end{small}

The user edits \texttt{vehicle.yaml} and \texttt{testcase.yaml}, runs
\texttt{vehron run -{-}case}, and gets a self-contained output package under
\texttt{veh-case-1/output/}. VEHRON prints a spec summary and progress during
the run; the package contains the timeseries, summary JSON, copied and
resolved inputs, and plots.

\section{Verification, Validation, and Software Quality}
The software quality and test-surface description in this section refers to the
VEHRON v0.2.2 release~\cite{vehronrepo}.
The repository has unit tests and integration tests. Unit tests cover battery
models, HVAC behaviour, slot interfaces, longitudinal dynamics, reducer,
analytical motor envelopes, motor-map handling, charging, configuration
loading, resource resolution, and runtime paths. Integration tests cover
end-to-end BEV runs, energy balance, reference benchmark ranges, regen
behaviour, and case-package generation. 62 tests pass on the current codebase.

The test suite establishes software consistency; physical fidelity still depends on comparison with measurement data. What it covers: schema-validated
configuration loading, deterministic case outputs for the documented workflow,
runnable public examples, and regression-style benchmarks for the active BEV
path.

Two regression reference points are defined in the current test suite: city
stop-start on the sedan archetype gives
8.00\,km, SoC~0.948, 1779\,Wh net with visible regen contribution. MIDC
P1$\times$4\,+\,P2 on Case~E gives 10.667\,km, SoC~0.958, 1285\,Wh net.
These numbers are used to catch software drift from one revision to the next.

The next step is straightforward: compare VEHRON against measured vehicle
data. The low-order models are calibrated for comparative studies such as cycle-level energy, thermal response, and subsystem loading. Teams with validated subsystem models can
substitute them through the slot interfaces without modifying the engine —
that is the intended path to higher fidelity.

\section{Reference Vehicle Configurations and Indicative Comparisons}
\label{sec:refvehicle}
Reference archetypes in VEHRON have fully stated parameters — every
assumption is in the YAML file, not buried in code. Six cases are presented
here: Cases~A--D are calibrated against WLTP, Cases~E--F against MIDC.
All are parameter sets assembled from publicly available technical data for
representative production
vehicles~\cite{hyundai2025ioniq6,tesla2024model3,bmw2024ix1,renault2024r5,mahindra2025be6,tata2024nexonev}.
They are reference configurations built from public data and stated assumptions. Parameters (mass, dimensions, drag, tyre class,
motor ratings, battery capacity) come from public domain sources. These
comparisons are included to show that VEHRON lands in the right operating
range. WLTP definitions and regulatory context are
in~\cite{unece2021r154,eu2024wltp}.

All numerical results in Tables~\ref{tab:wltp_benchmark}--\ref{tab:midc_benchmark}
and Figures~\ref{fig:wltp_motion_soc}--\ref{fig:midc_soc_comparison} refer to
VEHRON v0.2.2~\cite{vehronrepo}. The reported distances, SoC values, energy
budgets, and consumption figures are taken from the generated
\texttt{summary.json} files, with plots generated from the corresponding
\texttt{timeseries.csv} outputs.

Table~\ref{tab:model_index} lists all six cases with their parameter sources
and the cycle used for each comparison.

\begin{table}[h]
  \centering
  \caption{Reference parameter sets. Cases~A--D are calibrated against WLTP
           Class~3b; Cases~E--F against MIDC P1$\times$4\,+\,P2.
           Citations point to the public source used to assemble each case.}
  \label{tab:model_index}
  \footnotesize
  \setlength{\tabcolsep}{5pt}
  \begin{tabular}{@{}llllr@{}}
    \toprule
    Case & Description & Battery & Cycle & Source \\
    \midrule
    A & Compact fastback sedan, RWD  & 84\,kWh & WLTP & \cite{hyundai2025ioniq6} \\
    B & Large fastback sedan, RWD    & 82\,kWh & WLTP & \cite{tesla2024model3} \\
    C & Compact SUV/crossover, AWD   & 65\,kWh & WLTP & \cite{bmw2024ix1} \\
    D & City hatchback, FWD          & 52\,kWh & WLTP & \cite{renault2024r5} \\
    E & Compact SUV coupe, RWD       & 59\,kWh & MIDC & \cite{mahindra2025be6} \\
    F & Compact SUV, FWD             & 40.5\,kWh & MIDC & \cite{tata2024nexonev} \\
    \bottomrule
  \end{tabular}
\end{table}

\subsection{Configuration Methodology}

For each case, the following parameters are sourced from publicly available
technical data (brochures, EU configurators, press-release specifications) for
the named production vehicle and used as inputs to the VEHRON archetype YAML:
\begin{itemize}[leftmargin=1.5em]
  \item Published cycle energy consumption (kWh/100\,km) --- used as the
        calibration target, not as a guaranteed output
  \item Kerb weight and overall vehicle dimensions
  \item Drag coefficient $C_d$
  \item Tyre size (used to select $C_{rr}$ class)
  \item Motor peak power and peak torque
  \item Usable battery capacity
\end{itemize}
Quantities not directly available in public sources are estimated by standard
engineering practice and documented in the YAML header of each archetype.
The resulting configuration is a VEHRON case that resembles the named vehicle
in its key physical parameters; it is a reference vehicle assembled from public parameters.
Frontal area is taken as $A = \text{width} \times \text{height} \times k_f$,
where $k_f$ is a body-form factor (0.78 for streamlined fastback profiles,
0.80 for upright hatchback/coupe profiles, 0.84 for
crossover/SUV bodies). VEHRON uses these values as documented body-form
assumptions for reference-vehicle construction, with Hucho providing the
general aerodynamic basis for relating frontal area and body shape rather than
those exact tabulated constants~\cite{hucho1998aerodynamics}. Rolling resistance
coefficient $C_{rr}$ is assigned by tyre-class lookup (LRR EV tyres: 0.0065;
standard EV touring tyres: 0.0070--0.0075; SUV all-season:
0.0080). These are VEHRON engineering assumptions for representative tyre
classes; SAE~J2452 is cited for rolling-resistance measurement methodology
rather than as the source of these exact lookup values~\cite{sae_j2452}.
Transmission efficiency reflects whether a helical single-speed (0.985) or
smaller-unit single-speed (0.982) gearbox is used; these values are likewise
treated as documented engineering assumptions in the reference archetypes.

\textbf{Motor model.} All six cases use the same \texttt{analytical} motor
model. The only inputs that differ between cases are the nameplate
specifications (peak power, peak torque, maximum speed) and two scalar
calibration parameters: \texttt{base\_efficiency} and
\texttt{inverter\_efficiency}. The former reflects the motor technology class
inferred from public specifications (PMSM-class single motor: 0.93;
wound-rotor synchronous or dual-motor aggregate:
0.92)~\cite{agrawal2024traction}. The latter reflects the inverter technology
class (silicon carbide capable: 0.985; conventional silicon:
0.980)~\cite{shi2023sic}.

\textbf{Drivetrain efficiency.} All archetypes set
\texttt{drivetrain\_efficiency} to 1.0; drivetrain losses are fully
represented by the transmission, motor-efficiency, and inverter-efficiency
modules to avoid double-counting.

\textbf{Calibration of \texttt{regen\_blend\_factor}.} The parameter $\beta$
(Section~\ref{sec:regen}) is set per vehicle by energy-balance calibration
against the published cycle consumption figure. Fix all physical parameters,
run the cycle with $\beta = 1.0$, then adjust $\beta$ until $E_{\mathrm{net}}$
matches the published figure to within $\sim$1\%. The calibrated values
reflect known vehicle strategies:
\begin{itemize}[leftmargin=1.5em]
  \item Case~A (84\,kWh, $\beta = 0.83$): aggressive one-pedal regen; high
        capture fraction.
  \item Case~B (82\,kWh, $\beta = 0.61$): moderate standard regen mode; lower
        capture on gentle WLTP deceleration events.
  \item Cases~C and D (65 and 52\,kWh, $\beta = 0.72$): adjustable
        paddle-based regen at a moderate selection level.
  \item Case~E (59\,kWh, $\beta = 0.80$): aggressive i-Pedal one-pedal
        capable regen strategy.
  \item Case~F (40.5\,kWh, $\beta = 0.72$): four-level paddle regen at
        typical mixed setting.
\end{itemize}

\subsection{Analytical Energy Budget}

For each reference vehicle, the energy budget over the test cycle can be
decomposed analytically. Let $\eta_{\mathrm{chain}} = \eta_{\mathrm{trans}}
\times \eta_{\mathrm{motor}} \times \eta_{\mathrm{inv}}$ be the overall
drivetrain efficiency and $f_{\mathrm{regen}}$ the regen fraction
(blend~$\times$~regen efficiency). Then:
\begin{align}
  E_{\mathrm{aero}}    &= \int \tfrac{1}{2}\rho C_d A\,v^3\,\mathrm{d}t,\\
  E_{\mathrm{roll}}    &= \int m g C_{rr} v\,\mathrm{d}t,\\
  E_{\mathrm{wheel}}   &= E_{\mathrm{inertia}} + E_{\mathrm{aero}}
                          + E_{\mathrm{roll}},\\
  E_{\mathrm{drive}}   &= E_{\mathrm{wheel}} / \eta_{\mathrm{chain}},\\
  E_{\mathrm{regen}}   &= E_{\mathrm{inertia}} \times f_{\mathrm{regen}},\\
  E_{\mathrm{net}}     &= E_{\mathrm{drive}} - E_{\mathrm{regen}}
                          + E_{\mathrm{aux}} \cdot t_{\mathrm{cycle}}
                          + E_{\mathrm{hvac}}.
\end{align}
This decomposition is the basis for the header comments in each archetype YAML
file, and the archived reference-run \texttt{summary.json} outputs reproduce
these figures to within rounding.

\subsection{WLTP Results: Cases A--D}

Cases~A--D were run against the packaged WLTP Class~3b speed trace
(\texttt{wltp\_class3b.csv}, 1800~s, 23.27~km,
$v_{\max} = 131.3$~km/h). Table~\ref{tab:wltp_benchmark} summarises the
results alongside the published WLTP figure for the corresponding similar
production vehicle.

\begin{table}[h]
  \centering
  \caption{VEHRON WLTP comparison results for Cases~A--D.
           Gap = (VEHRON $-$ published) / published.}
  \label{tab:wltp_benchmark}
  \footnotesize
  \begin{tabular}{@{}lrrr@{}}
    \toprule
    Case & VEHRON & Published & Gap \\
         & (kWh/100\,km) & (kWh/100\,km) & (\%) \\
    \midrule
    A: compact fastback, 84\,kWh  & 13.21 & 13.5~\cite{hyundai2025ioniq6} & $-2.1$ \\
    B: large fastback, 82\,kWh    & 14.02 & 14.3~\cite{tesla2024model3}   & $-2.0$ \\
    C: compact SUV, 65\,kWh       & 17.81 & 18.0~\cite{bmw2024ix1}        & $-1.1$ \\
    D: city hatchback, 52\,kWh    & 13.78 & 13.9~\cite{renault2024r5}     & $-0.9$ \\
    \bottomrule
  \end{tabular}
\end{table}

All four cases fall within 2.5\% of the published figure. The small negative
gap indicates that these reference configurations land slightly below the
published WLTP values while remaining in the same operating range.
Table~\ref{tab:energy_budget} gives the full energy budget for each case.

\begin{sidewaystable}[p]
  \centering\small
  \caption{VEHRON energy budget for Cases~A--D over the WLTP Class~3b cycle
           (23.27~km). Road-load columns show wheel-side work; drivetrain
           losses account for the gap between wheel traction and
           $E_\text{drive}$. All quantities are written to
           \texttt{summary.json} in every case output.}
  \label{tab:energy_budget}
  \setlength{\tabcolsep}{7pt}
  \begin{tabular}{@{}lrrrrrrrrr@{}}
    \toprule
    Case & $E_\text{aero}$ & $E_\text{roll}$ & $E_\text{inertia}$
         & $E_\text{drive}$ & $E_\text{regen}$ & $E_\text{aux}$
         & $E_\text{hvac}$ & $E_\text{net}$ & $\beta$ \\
         & (Wh) & (Wh) & (Wh) & (Wh) & (Wh) & (Wh) & (Wh) & (Wh) & \\
    \midrule
    A: compact fastback, 84\,kWh &  933 & 1010 & 2037 & 3964 & 1045 & 100 & 53 & 3073 & 0.83 \\
    B: large fastback, 82\,kWh   &  978 &  983 & 1842 & 3779 &  670 & 100 & 53 & 3262 & 0.61 \\
    C: compact SUV, 65\,kWh      & 1343 & 1322 & 2167 & 4847 &  868 & 110 & 54 & 4143 & 0.72 \\
    D: city hatchback, 52\,kWh   & 1211 &  874 & 1529 & 3660 &  598 &  90 & 53 & 3205 & 0.72 \\
    \bottomrule
  \end{tabular}
\end{sidewaystable}

The road-load split varies meaningfully across the four cases. Case~C
(heavier SUV-profile, 2085\,kg) carries the highest rolling-resistance and
inertia work, giving the largest $E_{\mathrm{drive}}$. Case~A achieves the
best regen recovery fraction (26\% of drive energy) owing to its high
$\beta = 0.83$; Case~B recovers the least (18\%) despite similar aerodynamics,
because its lower $\beta = 0.61$ leaves more kinetic energy to friction braking.

\begin{figure}[t]
  \centering
  \includegraphics[width=\linewidth]{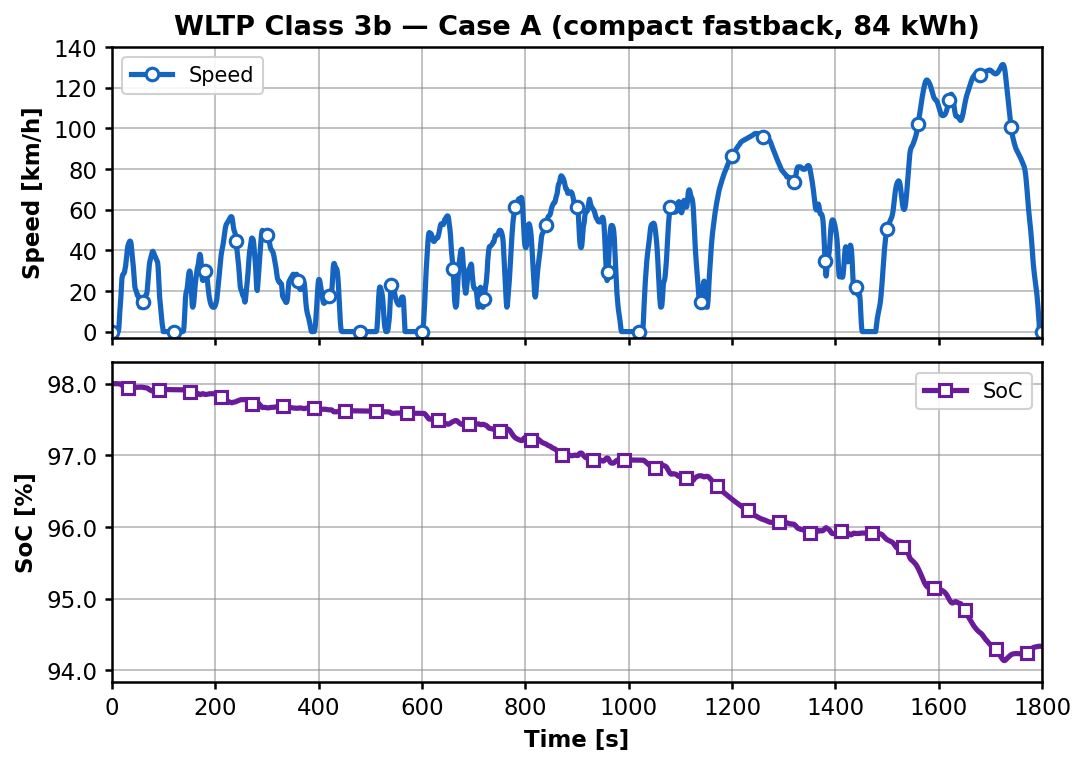}
  \caption{Motion and SoC response for Case~A (compact fastback, 84\,kWh)
           on the WLTP Class~3b speed trace. VEHRON produces
           13.21\,kWh/100\,km; the published figure for a similarly specified
           production vehicle is 13.5\,kWh/100\,km ($-2.1\%$).}
  \label{fig:wltp_motion_soc}
\end{figure}

\begin{figure}[t]
  \centering
  \includegraphics[width=\linewidth]{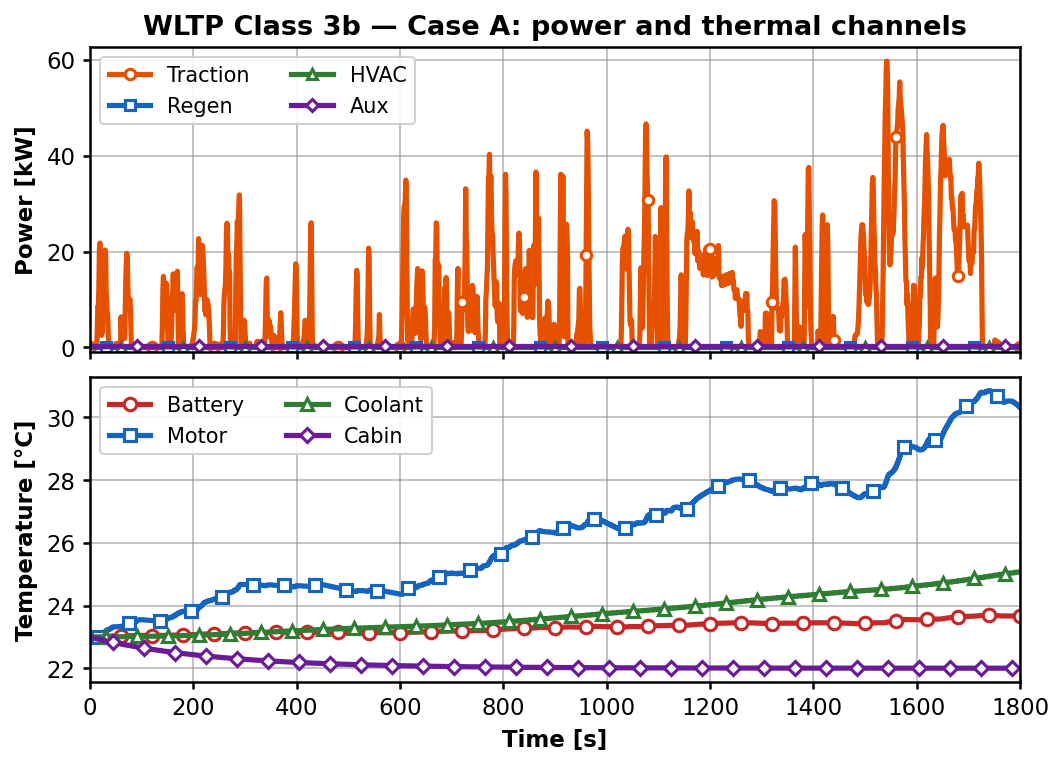}
  \caption{Power and thermal response for Case~A on the WLTP Class~3b speed
           trace, showing principal electrical power channels and the coupled
           battery, motor, coolant, and cabin temperatures.}
  \label{fig:wltp_power_thermal}
\end{figure}

Figures~\ref{fig:wltp_motion_soc} and~\ref{fig:wltp_power_thermal} show the
VEHRON response for Case~A — the compact fastback with the highest regen
fraction. Case~A ends at 13.21\,kWh/100\,km against a published figure of
13.5\,kWh/100\,km, which shows the reference configuration behaves like a
plausible passenger EV on this cycle.

Figures~\ref{fig:wltp_motion_soc} and~\ref{fig:wltp_power_thermal} together
show a transient operating history. The regen contribution is visible in the
power channel during deceleration events, consistent with Case~A's
$\beta = 0.83$.

\subsection{MIDC Results: Cases E--F}

Cases~E and~F were run against the packaged MIDC P1$\times$4\,+\,P2 speed
trace (\texttt{midc\_p1x4\_p2.csv}, 1180~s, 10.67~km, $v_{\max} = 90$~km/h).
The cycle follows CMVR Annexure~IV-B~\cite{cmvr_annexure_ivb}, with Part~I
(195\,s elementary urban) repeated four times and Part~II (400\,s extra-urban)
appended, as specified for M1/M2 electric-vehicle range measurement in
AIS-040~\cite{ais040rev1_2015,ais040rev1_amd2024}. The P1 segment runs
$t = 0$--780\,s; P2 runs $t = 781$--1180\,s.
Table~\ref{tab:midc_benchmark} summarises the results.

\begin{table}[h]
  \centering
  \caption{VEHRON MIDC comparison results for Cases~E--F alongside
           ARAI-certified claimed ranges. Gap = (VEHRON $-$ implied
           ARAI consumption) / implied ARAI consumption.}
  \label{tab:midc_benchmark}
  \scriptsize
  \setlength{\tabcolsep}{4pt}
  \begin{tabular}{@{}lrrrr@{}}
    \toprule
    Case & VEHRON & ARAI claimed & Implied ARAI & Gap \\
         & (kWh/100\,km) & range (km) & (kWh/100\,km) & (\%) \\
    \midrule
    E: SUV coupe, 59\,kWh  & 12.05 & 535~\cite{mahindra2025be6} & 11.03 & $+9.2$ \\
    F: compact SUV, 40.5\,kWh & 13.37 & 465~\cite{tata2024nexonev} & 8.71  & $+53.5$ \\
    \bottomrule
  \end{tabular}
\end{table}

Case~E sits within 10\% of the implied ARAI consumption. Case~F shows a larger
gap: its ARAI-implied consumption is notably lower than the VEHRON result for
the corresponding reference configuration, making it the least aligned of the
two MIDC cases in this set.

\begin{figure}[t]
  \centering
  \includegraphics[width=\linewidth]{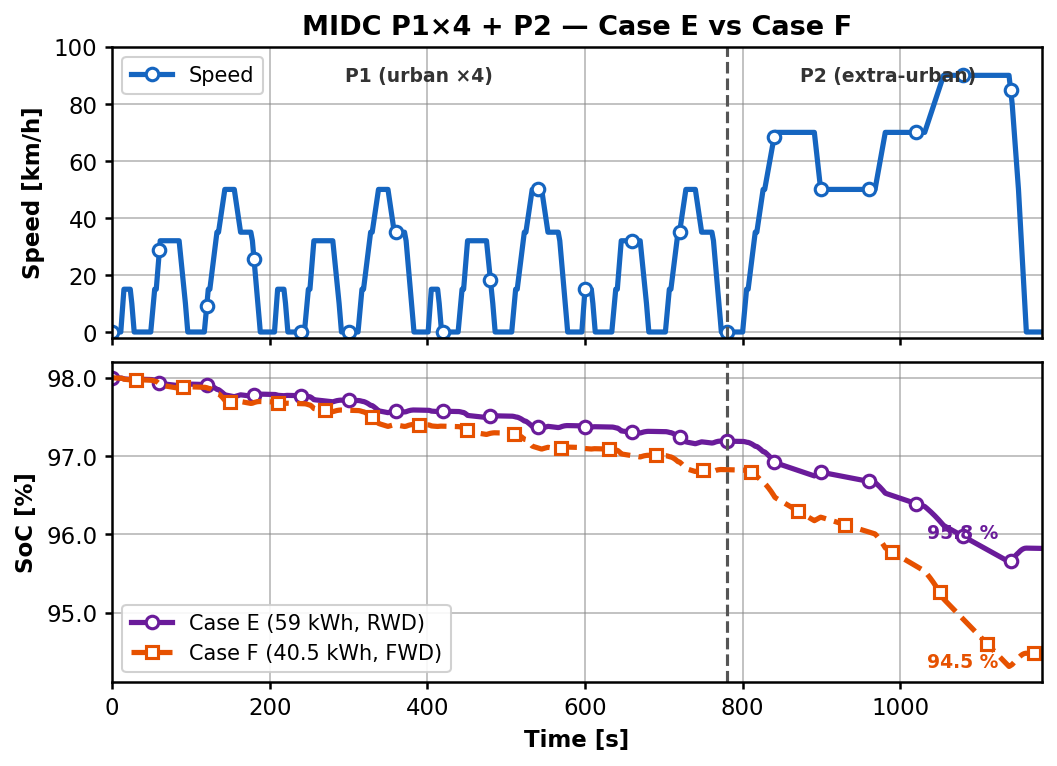}
  \caption{MIDC P1$\times$4\,+\,P2 (1180\,s, 10.67\,km): speed trace and
           SoC response for Case~E (59\,kWh RWD, solid) and Case~F
           (40.5\,kWh FWD, dashed). Dashed vertical line marks the P1/P2
           phase boundary at $t = 780$\,s. Case~E ends at SoC\,=\,95.8\%;
           Case~F at 94.5\%.}
  \label{fig:midc_soc_comparison}
\end{figure}

Figure~\ref{fig:midc_soc_comparison} shows the SoC response for both cases
across the MIDC cycle. The P1/P2 phase boundary is marked; the larger SoC
drop for Case~F in the P2 extra-urban segment reflects its higher aerodynamic
drag and conventional tyre compound relative to Case~E.

If a study needs closer platform fidelity, calibrate the YAML parameters
against measured data and replace the subsystem models through the external
battery and HVAC slots.

\section{Availability, Installation, and Citation}
VEHRON~\cite{vehronrepo} is available as open-source software under the GNU
Affero General Public License, version 3 only (AGPL-3.0-only). The public
project homepage is \url{https://vehron.org}, and the public source repository
is \url{https://github.com/vehron-dev/vehron}. Release
citation metadata in the repository points users to the archived software
release under DOI \url{https://doi.org/10.5281/zenodo.19820111} for version
0.2.2.

VEHRON is released under AGPL-3.0-only. Architecturally, the external battery
and HVAC slots allow proprietary subsystem models to remain outside the VEHRON
repository while still being used in the vehicle-level workflow. That is a
software-architecture statement rather than legal advice; users integrating
proprietary code should review their own licensing obligations.

Installation is \texttt{pip install .} from a checkout, or \texttt{pip install
-e .} for development. Archetypes, testcases, and drive-cycle CSVs are
bundled as package data. A PyPI release is planned once the public API is more
stable.

Cite the archived release corresponding to the version used, per the
\texttt{CITATION.cff} in the repository.

\section{Limitations and Roadmap}
The active path is BEV longitudinal simulation. This paper focuses on that
path; other components in the repository are still evolving.

The models are low-order. The battery, cabin, motor, and coolant models are simplified low-order models for comparative studies such as cycle-level energy, thermal response, and subsystem loading, not for certifying a pack design or
sizing a cooling system. The AC charging model is a crude CC/CV approximation.
The motor efficiency map uses nearest-neighbour lookup. Route input is
parametric or speed-trace CSV; GPS replay and elevation profiles are not yet
supported.

Near-term priorities: tighter validation against reference data, stronger
tests around slot interfaces and output contracts, and cleaner release-level
separation between supported features and experimental ones. The roadmap beyond
that includes hybrid-electric and fuel-cell powertrain paths, richer route
inputs, and a broader validated archetype library.

\section{Conclusion}
VEHRON~\cite{vehronrepo} contributes a structured simulation workflow built
around YAML-defined vehicles, testcases, and case artifacts. YAML-stated
vehicle and testcase definitions, a clean subsystem slot architecture, and
deterministic case packaging — together, they make BEV studies easier to
rerun, compare, and inspect.

The current scope is deliberately narrow: BEV longitudinal simulation, tested,
with documented extension points for battery and HVAC.
The next steps are broader validation against measured data, tighter slot
interfaces, and expansion toward hybrid and fuel-cell powertrain paths.
The foundation is in place.


\begin{thebibliography}{9}

\bibitem{vehronrepo}
S.~Natarajan.
\newblock \emph{VEHRON}.
\newblock Version 0.2.2, 2026.
\newblock Software available at \url{https://github.com/vehron-dev/vehron}.
\newblock Archived release DOI: \url{https://doi.org/10.5281/zenodo.19820111}.

\bibitem{advisor2002}
T.~Markel, A.~Brooker, T.~Hendricks, V.~Johnson, K.~Kelly, B.~Kramer,
M.~O'Keefe, S.~Sprik, and K.~Wipke.
\newblock ADVISOR: a systems analysis tool for advanced vehicle modeling.
\newblock \emph{Journal of Power Sources}, 110(2):255--266, 2002.
\newblock doi:10.1016/S0378-7753(02)00163-2.

\bibitem{autonomie2011}
Argonne National Laboratory.
\newblock \emph{Autonomie Vehicle System Simulation Tool}.
\newblock 2026.
\newblock \url{https://www.anl.gov/taps/autonomie-vehicle-system-simulation-tool}.
\newblock Accessed April 27, 2026.

\bibitem{fastsim2018}
A.~Brooker, J.~Gonder, L.~Wang, E.~Wood, S.~Lopp, and L.~Ramroth.
\newblock FASTSim: A model to estimate vehicle efficiency, cost and performance.
\newblock SAE Technical Paper 2015-01-0973, 2015.
\newblock doi:10.4271/2015-01-0973.
\newblock NREL publication page:
\href{https://research-hub.nrel.gov/en/publications/fastsim-a-model-to-estimate-vehicle-efficiency-cost-and-performan-2/}{FASTSim publication record}.

\bibitem{openmodelica2020}
P.~Fritzson, A.~Pop, K.~Abdelhak, A.~Asghar, B.~Bachmann, W.~Braun,
D.~Bouskela, R.~Braun, L.~Buffoni, and N.~Casella et al.
\newblock The OpenModelica integrated environment for modeling, simulation, and
model-based development.
\newblock \emph{Modeling, Identification and Control}, 41(4):241--295, 2020.
\newblock doi:10.4173/mic.2020.4.1.

\bibitem{gillespie1992fvd}
T.~D.~Gillespie.
\newblock \emph{Fundamentals of Vehicle Dynamics}.
\newblock SAE International, Warrendale, PA, 1992.
\newblock doi:10.4271/R-114.
\newblock \url{https://saemobilus.sae.org/books/fundamentals-vehicle-dynamics-r-114}.

\bibitem{rajamani2012vdc}
R.~Rajamani.
\newblock \emph{Vehicle Dynamics and Control}.
\newblock 2nd edition, Springer, 2012.
\newblock doi:10.1007/978-1-4614-1433-9.

\bibitem{liu2013hev}
W.~Liu.
\newblock \emph{Introduction to Hybrid Vehicle System Modeling and Control}.
\newblock John Wiley \& Sons, 2013.

\bibitem{hyundai2025ioniq6}
Hyundai Motor Europe.
\newblock \emph{IONIQ 6 Brochure}.
\newblock 2025.
\newblock \url{https://dmassets.hyundai.com/is/content/hyundaiautoever/Hyundai_IONIQ-6_Brochure-EN_20251202_1280x720_40ppdf}.
\newblock Accessed April 2026.

\bibitem{tesla2024model3}
Tesla, Inc.
\newblock \emph{Model 3 EU Technical Specifications}.
\newblock 2024.
\newblock \url{https://www.tesla.com/en_eu/model3}.
\newblock Accessed April 2026.

\bibitem{bmw2024ix1}
BMW Group.
\newblock \emph{BMW iX1 xDrive30 Technical Data}.
\newblock 2024.
\newblock \url{https://www.bmw.com/en/models/ix1/xdrive30.html}.
\newblock Accessed April 2026.

\bibitem{renault2024r5}
Renault Group.
\newblock \emph{Renault 5 E-Tech 100\% Electric Technical Specifications}.
\newblock 2024.
\newblock \url{https://www.renault.co.uk/electric-vehicles/r5-e-tech-electric/specifications.html}.
\newblock Accessed April 2026.

\bibitem{unece2021r154}
United Nations Economic Commission for Europe.
\newblock \emph{UN Regulation No. 154 - Worldwide harmonized Light vehicles Test Procedure (WLTP)}.
\newblock 2021.
\newblock \url{https://unece.org/transport/documents/2021/02/standards/un-regulation-no-154-worldwide-harmonized-light-vehicles-test}.
\newblock Accessed April 24, 2026.

\bibitem{eu2024wltp}
Publications Office of the European Union.
\newblock \emph{Worldwide harmonised Light-duty vehicles Test Procedure (WLTP) and Real Driving Emissions (RDE)}.
\newblock 2024.
\newblock \url{https://op.europa.eu/en/publication-detail/-/publication/12790537-f4f0-11e9-8c1f-01aa75ed71a1/language-en}.
\newblock Accessed April 24, 2026.

\bibitem{hu2012ecm}
X.~Hu, S.~E.~Li, and H.~Peng.
\newblock A comparative study of equivalent circuit models for Li-ion
batteries.
\newblock \emph{Journal of Power Sources}, 198:359--367, 2012.
\newblock doi:10.1016/j.jpowsour.2011.10.013.

\bibitem{su2019ecm}
J.~Su, M.~Lin, S.~Wang, J.~Li, J.~Coffie-Ken, and F.~Xie.
\newblock An equivalent circuit model analysis for the lithium-ion battery pack
in pure electric vehicles.
\newblock \emph{Measurement and Control}, 52(3--4):193--201, 2019.
\newblock doi:10.1177/0020294019827338.

\bibitem{marcos2014cabin}
D.~Marcos, F.~J.~Pino, C.~Bordons, and J.~J.~Guerra.
\newblock The development and validation of a thermal model for the cabin of a
vehicle.
\newblock \emph{Applied Thermal Engineering}, 66(1--2):646--656, 2014.
\newblock doi:10.1016/j.applthermaleng.2014.02.054.

\bibitem{torregrosa2015cabin}
B.~Torregrosa-Jaime, F.~Bjurling, J.~M.~Corberan, and F.~Di Sciullo.
\newblock Transient thermal model of a vehicle's cabin validated under variable
ambient conditions.
\newblock \emph{Applied Thermal Engineering}, 75:45--53, 2015.
\newblock doi:10.1016/j.applthermaleng.2014.05.074.

\bibitem{noreen2019thermal}
G.~J.~Marshall, C.~P.~Mahony, M.~J.~Rhodes, S.~R.~Daniewicz, N.~Tsolas, and
S.~M.~Thompson.
\newblock Thermal management of vehicle cabins, external surfaces, and onboard
electronics: An overview.
\newblock \emph{Engineering}, 5(5):954--969, 2019.
\newblock doi:10.1016/j.eng.2019.02.009.

\bibitem{cabreview2024}
L.~Zhao, Q.~Zhou, and Z.~Wang.
\newblock A systematic review on modelling the thermal environment of vehicle
cabins.
\newblock \emph{Applied Thermal Engineering}, 257:124142, 2024.
\newblock doi:10.1016/j.applthermaleng.2024.124142.

\bibitem{agrawal2024traction}
S.~Agrawal and A.~Mistry.
\newblock Review of traction motors for electric vehicle application.
\newblock \emph{Journal of Electrical Systems}, 20(3):2737--2747, 2024.
\newblock \url{https://journal.esrgroups.org/jes/article/download/4522/3325/8206}.

\bibitem{shi2023sic}
B.~Shi, A.~I.~Ramones, Y.~Liu, H.~Wang, Y.~Li, S.~Pischinger, and J.~Andert.
\newblock A review of silicon carbide {MOSFETs} in electrified vehicles:
  Application, challenges, and future development.
\newblock \emph{IET Power Electronics}, 2023.
\newblock doi:10.1049/pel2.12524.

\bibitem{sae_j2452}
{SAE International}.
\newblock \emph{Stepwise Coastdown Methodology for Measuring Tire Rolling
  Resistance}.
\newblock Surface Vehicle Recommended Practice J2452. SAE International, 2017.

\bibitem{szumska2025regen}
E.~M.~Szumska.
\newblock Regenerative braking systems in electric vehicles: A comprehensive
  review of design, control strategies, and efficiency challenges.
\newblock \emph{Energies}, 18(10):2422, 2025.
\newblock doi:10.3390/en18102422.

\bibitem{hucho1998aerodynamics}
W.-H.~Hucho.
\newblock \emph{Aerodynamics of Road Vehicles}.
\newblock 4th edition. SAE International, 1998.
\newblock ISBN 978-0768000290.

\bibitem{ais040rev1_2015}
Automotive Industry Standards Committee / Automotive Research Association of India.
\newblock \emph{AIS-040 (Rev.1): 2015 --- Electric Power Train Vehicles ---
  Method of Measuring the Range}.
\newblock ARAI, Pune, February 2015.

\bibitem{ais040rev1_amd2024}
Automotive Industry Standards Committee / Automotive Research Association of India.
\newblock \emph{Amendment No. 3 (04/2024) to AIS-040 (Rev.1): 2015 ---
  Electric Power Train Vehicles --- Method of Measuring the Range}.
\newblock ARAI, Pune, April 2024.
\newblock Amendment~3 specifies MIDC for M1/M2 as Part~I (four elementary urban
  cycles of 195\,s each) plus Part~II.

\bibitem{cmvr_annexure_ivb}
{Ministry of Road Transport and Highways, Government of India}.
\newblock \emph{The Central Motor Vehicles Rules, 1989, Annexure~IV-B:
  Driving Cycles and Cold Start}.
\newblock MoRTH, 1989 (as amended).
\newblock Part~I and Part~II MIDC speed tables at pp.~626--627 of the
  machine-readable mirror:
\url{https://ebook.commerciallawpublishers.com/fa/cmvr/files/basic-html/page626.html}
and
\url{https://ebook.commerciallawpublishers.com/fa/cmvr/files/basic-html/page627.html}.
\newblock Accessed 2026-04-27.

\bibitem{mahindra2025be6}
Mahindra \& Mahindra Ltd.
\newblock \emph{BE 6 Technical Specifications}.
\newblock 2025.
\newblock Official brochure:
\url{https://www.mahindraelectricsuv.com/on/demandware.static/-/Library-Sites-eSUVSharedLibrary/default/dwc19a84f6/Be-6e/5050-MAH-BE-6-Brochure-v18-AW-150DPI.pdf}.
\newblock Accessed April 2026.

\bibitem{tata2024nexonev}
Tata Motors Ltd.
\newblock \emph{Nexon EV Long Range Technical Specifications}.
\newblock 2024.
\newblock \url{https://ev.tatamotors.com/content/tml/ev/in/en/nexon/ev/specifications.html}.
\newblock Accessed April 2026.

\bibitem{gtsuite_web}
{Gamma Technologies LLC}.
\newblock {GT-SUITE} vehicle simulation software.
\newblock \url{https://www.gtisoft.com/gt-suite/}, 2024.
\newblock Accessed April 2026.

\bibitem{ipg_carmaker_web}
{IPG Automotive GmbH}.
\newblock {CarMaker} vehicle simulation.
\newblock \url{https://ipg-automotive.com/en/products-solutions/software/carmaker/}, 2024.
\newblock Accessed April 2026.

\end{thebibliography}
\end{document}